\documentclass[a4paper,11pt]{article}
\usepackage{pos}
\usepackage{multicol}
\usepackage{amssymb,amsmath}
\usepackage{url}
\usepackage{enumitem}
\title{Open Science in Lattice Gauge Theory community}
\newcommand{\be}{\begin{equation}}
\newcommand{\ee}{\end{equation}}
\newcommand{\bea}{\begin{eqnarray}}
\newcommand{\eea}{\end{eqnarray}}

\author*[a]{Andreas Athenodorou}
\author[b]{Ed Bennett}
\author[b]{Julian Lenz}
\author[c]{Elli Papadopoullou}

\affiliation[a]{Computation-based Science and Technology Research Center, The Cyprus Institute, Cyprus}
\affiliation[b]{Swansea Academy of Advanced Computing, Swansea University, Fabian Way, Swansea SA1 8EN, UK}
\affiliation[c]{ATHENA research center, Artemidos 6 \& Epidavrou 15125, Marousi, Greece}

\emailAdd{a.athenodorou@cyi.ac.cy}
\emailAdd{e.j.bennett@swansea.ac.uk}
\emailAdd{j.j.lenz@swansea.ac.uk}
\emailAdd{elli.p@athenarc.gr}

\abstract{Open science aims to make scientific research processes, tools and results accessible to all scientific communities, creating trust in science and enabling digital competences to be realized in research, leading to increased innovation. It provides standard and transparent pathways to conducting research and fosters best practices for collecting, analysing, preserving, sharing and reusing data, software, workflows and other outputs through collaborative networks. Open Science appears to be becoming the norm with its applications spanning throughout the whole research cycle of a project. The importance of making Open Science a reality is nowadays reflected in funding policies, research infrastructure and politics. In these proceedings we present the basic Open Science principles explaining briefly best practices for materialising Open Science. Subsequently, we present the results of the landscaping survey of Open Science in the Lattice Gauge Theories community. Finally, we provide directions in which the Lattice Gauge Theory community could move in order to enhance Openness and FAIRness (Findability, Accessibility, Interoperability, Reusability) in Science.}

\FullConference{%
  The 39th International Symposium on Lattice Field Theory (Lattice2022),\\
  8-13 August, 2022 \\
  Bonn, Germany 
}

\begin{document}
\maketitle

% *********************************************
% * Introduction
% *********************************************
\section{Introduction}
\label{sec:introduction}
\vspace{-0.25cm}
Open Science is the approach of making scientific research at different points in a project timeline---including (but not limited to) publications, data, and software---accessible to all, as well as its subsequent dissemination. Open science is achieved by transparent practices that enhance inter- and multi-disciplinary collaborative networks. Such practices include the implementation of the principles that data should be Findable, Accessible, Interoperable, and Reusable (FAIR).

Open Science is a concept consisting of several aspects, with a taxonomy that has a broad range of subjects~\cite{pontika2015open}. One of the most common concepts that we have all encountered at some point is Open Access. Open Access encompasses making outputs openly available; commonly-used routes include ``Gold''---publishing in an open-access journal---and ``Green''---where one self-archives a version of the peer-reviewed article in an openly accessible archive, such as \href{https://arxiv.org/}{\texttt{arXiv}}.

For most modern scientific research, however, publishing an article in an open access journal is not enough to meet the Open Science requirements. The absolute goal of Open Science is to enable reusability and reproducibility of scientific outputs. One should therefore provide access to full raw data, as well as software and documentation which could help others reproduce the final results. Open Science is increasingly required by funding bodies, such as Horizon Europe~\cite{doi/10.2777/69533}. To support the uptake of Open Science and improve the quality, efficiency, and responsiveness of research, the European Commission develops the European Open Science Cloud (EOSC)~\cite{doi/10.2777/940154}. EOSC is based on the federation of services for research, and has been initiated to ensure that European as well as international scientists enjoy the full benefits of data-driven science. It is co-designed and -developed with major infrastructures for research across Europe, such as \href{https://www.egi.eu/egi-for-eosc/}{EGI}, \href{https://connect.geant.org/tag/eosc}{GEANT}, and \href{https://www.openaire.eu/openaire-and-eosc}{OpenAIRE}, aiming to offer researchers and professionals in science and technology a virtual environment with free at the point of use, open and seamless services for storage, management, analysis and reuse of research data, across borders and scientific disciplines.

The concept of Open Science has a number of advantages which can, indeed, leverage the whole scientific lifecycle. Below, we outline a number of these advantages for research, researchers and the society at large.

From the point of view of research, Open Science secures the preservation of the intellectual legacy of research institutions while preventing any loss of data and the produced knowledge. Furthermore, it publishes scientific results in a timely manner, thus accelerating new discoveries. It also  enhances the citability and reusability of research results by assigning permissive and attribution licenses, and serves validation through reproducible practices, so that the final product can be verified.

Researchers following Open Science practices increase their visibility, as they are able to reach wider audiences by publishing in venues that are not paywalled and that support sharing of outputs that go beyond publications. They are, therefore, more likely to get cited by other researchers and gain credit for their intellectual work.

Moreover, Open Science has an important societal dimension. It builds on transparent processes to strengthen public trust in research and eliminate---to the extent possible---fake news, while reducing fear in scientific and technological advancements. 

But, practising Open Science is a time consuming procedure that requires a lot of effort from researchers, especially when dedicated support from professionals, like data stewards or data curators, is not provided. That means that, in order to facilitate Open Science in each scientific domain, it is essential to build competences along with collaborating to create best practices and tools that address the complexities of the given research community. This is the direction that will enable lattice gauge theorists to harmonize their research in accordance to Open Science principles.

In this work, we provide a first landscape of the current Open Science scenery in the field of Lattice Gauge Theory. The results enabled us to identify both the weaknesses and the challenges of this community to practicing Open Science. They can then feed efforts for the creation of best practices and tools to support the Lattice Gauge Theory researchers in applying Open and FAIR principles in their research. This contribution is organized in the following way: We first provide a description of the FAIR principles, and we then present our findings from the survey that was performed within the lattice community.
\vspace{-0.25cm}
\section{FAIR data and Open Science}
\label{sec:FAIR}
\vspace{-0.25cm}
The FAIR principles were coined as a term at the \href{https://www.lorentzcenter.nl/jointly-designing-a-data-fairport.html}{Lorentz} workshop in 2014. FAIR stands for Findable, Accessible, Interoperable and Reusable. One of the grand challenges defined in the workshop was the initiative of a data-intensive science for the facilitation of knowledge discovery, which should assist humans and machines to discover, access, integrate and analyse, task-appropriate data and their associated algorithms and workflows. These principles were subsequently published in 2016~\cite{wilkinson2016fair}. The authors intended to provide guidelines to improve the findability, accessibility, interoperability, and reusability of digital assets. The principles emphasise machine-actionability (i.e., the capacity of computational systems to find, access, interoperate, and reuse data with no or minimal human intervention), because humans increasingly rely on computational support to deal with data as a result of the increase in volume, complexity, and rapid production of data.

Fifteen principles and a set of fourteen metrics have been defined to quantify levels of FAIRness. The principles below refer to three types of entities: data (or any digital object), metadata (information about that digital object), and infrastructure.
\vspace{-0.25cm}
\subsection{Findable}
\label{sec:findable}
\vspace{-0.25cm}
The first step in (re)using data is to be able to find them. Data and metadata should be 
easily findable by both humans and machines (computers). Machine-readable metadata are essential for automatic discovery of datasets and services, so this is an essential component of the 
FAIRification process.
\begin{itemize}[noitemsep,topsep=1pt]
    \item (Meta)data are assigned a globally unique and persistent identifier (PIDs).
    \item Data are described with rich metadata.
    \item Metadata clearly and explicitly include the identifier of the data they describe.
    \item (Meta)data are registered or indexed in a searchable resource.
\end{itemize}
\noindent
To achieve findability of data one should make use of standard identification mechanisms that are persistent and unique, such as Digital Object Identifiers (DOIs). In addition one should provide a description of the naming conventions used to name the data files, to outline the approach towards search keywords and versioning, and specify the standards followed by metadata. The lattice community suffers with a lack of metadata standards; while the \href{https://www2.ccs.tsukuba.ac.jp/ILDG/}{ILDG}~\cite{ILDG_Karsch} metadata schema exists for the characterisation of SU(3) gauge configurations and correlators, there is a need for more general metadata schemas capable of describing all different kinds of lattice datasets.
\vspace{-0.25cm}
\subsection{Accessible}
\label{sec:accessible}
\vspace{-0.25cm}
Once the user finds the required data, they need to know how they can be accessed, possibly by including authentication and authorization. This requires that:
\begin{itemize}[noitemsep,topsep=1pt]
\item (Meta)data are retrievable by their identifier using a standardized communications protocol:
\begin{itemize}[noitemsep,topsep=1pt]
\item The protocol is open, free, and universally implementable.
\item The protocol allows for an authentication and authorization procedure.
\end{itemize}
\item Metadata are accessible, even when the data are no longer available.
\end{itemize}
\noindent
Good practises to ensure accessibility include the specification of which data will be made openly available. In case some data is kept closed one should provide rationale for doing so. The way data will be made available should be described. In addition it is essential to describe the methods or software tools needed to access the data. It would help if the documentation about the software needed to access the data is included. It would also be good to include the relevant software (e.g.~in open source code). One should specify where the data and associated metadata, documentation and code are deposited. In case there are any restrictions one should specify how access will be provided.
\vspace{-0.25cm}
\subsection{Interoperable}
\label{sec:interoperable}
\vspace{-0.25cm}
Data usually need to be integrated with other data. In addition, data needs to interoperate with applications or workflows for analysis, storage, and processing.
\begin{itemize}[noitemsep,topsep=1pt]

    \item (Meta)data use a formal, accessible, shared, and broadly applicable language for knowledge representation.
    
    \item(Meta)data use vocabularies that follow FAIR principles.
    
    \item(Meta)data include qualified references to other (meta)data.

\end{itemize}
\noindent
Good practises to ensure interoperability include to assess the interoperability of your data. One should specify the data and metadata vocabularies, standards or methodologies  which are needed to be followed to facilitate interoperability. In addition it helps to specify whether they will be using standard vocabulary for all data types present in the data set so that they allow inter-disciplinary interoperability. If not, it helps to provide mapping to more commonly used ontologies. Ontology is a description of the semantics of the data, providing a uniform way to enable communication by which different parties can understand each other.

\subsection{Reusable}
\label{sec:reusable}
\vspace{-0.25cm}
The ultimate goal of FAIR principles is to optimize the reuse of data. To achieve this, metadata and data should be well-described so that they can be replicated and/or combined in different settings.
\begin{itemize}[noitemsep,topsep=1pt]
    \item (Meta)data are richly described with a plurality of accurate and relevant attributes:
    \begin{itemize}[noitemsep,topsep=1pt]
    \item (Meta)data are released with a clear and accessible data usage license.
    
    \item (Meta)data are associated with detailed provenance.
    
    \item (Meta)data meet domain-relevant community standards.
\end{itemize}
\end{itemize}

\noindent
Good practises to achieve reusability should include the specification of how data will be licenced to permit the widest reuse possible. In addition one should specify when the data will be made available for reuse. If applicable, one should also explain why and for what period data embargo is needed. Furthermore one should also defined whether the data produced and/or used in the project is useable by third parties, in particular after the end of the project. If the reuse of some data is restricted, it would help to explain why this is so. An important aspect of reusability, is the description of how data quality assurance is ensured. Finally, one should identify the time for which the data will remain reusable.

\subsection{FAIR is not Open}
\label{sec:FAIRvsOpen}
\vspace{-0.25cm}
At this point, we should emphasize that {\bf FAIR data is not equal to Open data} and neither is a subset of the other. Although they overlap in the ``A'' (=Accessibility) in the definition of FAIR, practically, if a dataset is Open but not FAIR, it is still useless. Conversely, the Accessibility in FAIR is subject to well defined conditions and explicitly and deliberately does not address moral and ethical issues pertaining to the Openness of data.\footnote{In principle there could be good enough reasons to protect your data and services. These may include personal privacy, national security, commercialisation and competitiveness.} In the FAIR environment, the degree to which any dataset accessible is solely at the discretion of its owner. None of the FAIR principles necessitate data being Open, and the FAIR principles are still valuable for data which may never be made public, as internal users of the data still need to be able to find it, access it, interoperate it with other data, and make use of it.
\vspace{-0.25cm}

% *********************************************
% * Survey Results
% *********************************************
\section{Survey Results}
\label{sec:survey_results}
\vspace{-0.25cm}
The survey was announced on the ``Lattice News'' mailing list on the $20^{\rm th}$ of June 2022 and the results which are compiled within this manuscript have been extracted from the responses received by the start of the Lattice 2022 conference. In total 39 fully filled as well as 67 partially filled surveys had been received. The survey definition, raw results obtained, and the analysis workflow to generate the plots presented here can be found in Ref.~\cite{andreas_athenodorou_2022_6980070}.

\begin{figure}[h]
    \vspace{-0.7cm}
    \centering
    \includegraphics[height=4.05cm]{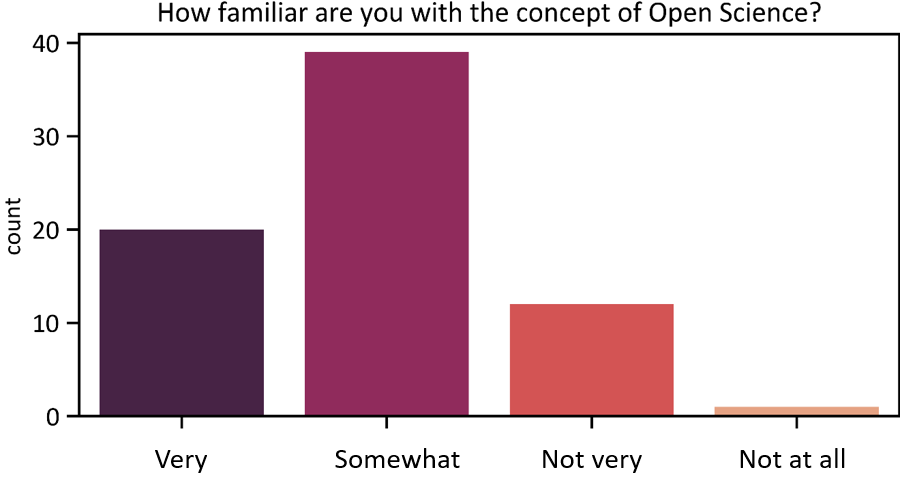}  \includegraphics[height=4.05cm]{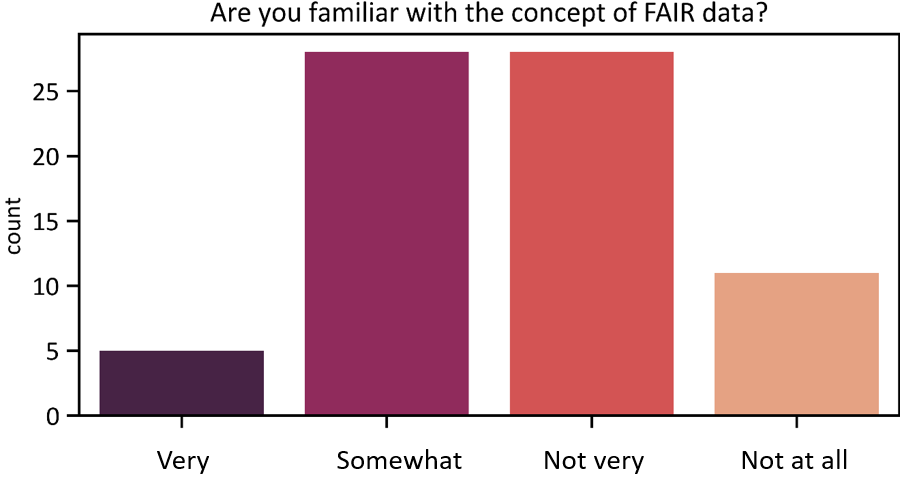} 
    \vspace{-0.25cm}
    \caption{ \underline{Left panel:} The results on the question ``How familiar are you with Open Science?''
    \underline{Right panel:} The results on the question ``How familiar are you with FAIR data?''}
    \label{fig:plots_familiarity_OS_FAIR}
    \vspace{-0.35cm}
\end{figure}

As a first step we investigated the familiarity of the Lattice community with the Open and FAIR concepts. For the question about how familiar the participant is with the concept of Open Science and FAIR data, the results are presented as histograms in Fig.~\ref{fig:plots_familiarity_OS_FAIR}. Clearly, most participants are familiar with Open Science. This is somewhat expected given that the concept of Open Science, especially the aspect of Open Access has been a subject of interest for quite a long time. However, when it comes to the familiarity with the concept of FAIR data it appears that the majority of survey participants know little about it.   

The way data are curated within the life-cycle of a research project should be documented in the so called Data Management Plan (DMP). A DMP is a formal document that outlines the ways in which data are collected, generated and/or processed throughout the lifespan of a research project, as well as after the project is completed, and therefore it should be tailored to the specific characteristics and needs of every project. A DMP is usually a deliverable as well as a living document that is continually edited and updated. A DMP is not a research assessment method and does not reflect the scientific quality of a project. Nevertheless, a DMP is typically a requirement of funding bodies both at proposal stage as well as during the implementation of all projects. It is one of the evaluated elements and if not properly designed, can lead to lower scoring of proposals. 

Good practises of handling Open data require the creation of a well documented DMP. An example of a template one can use to create a DMP is provided in the \href{https://ec.europa.eu/research/participants/data/ref/h2020/other/gm/reporting/h2020-tpl-oa-data-mgt-plan-annotated_en.pdf}{Horizon 2020 DMP template}. Furthermore, the author can use the \href{https://argos.openaire.eu/splash/}{Argos} tool developed by OpenAIRE to assist the creation of a well documented and complete DMP. When asked whether they have ever created a Data Management Plan for their research, $\sim 40 \%$ of respondents replied positively while  $\sim 60 \%$ negative. Thus, effort should be invested in creating awareness on DMPs as well as practical examples of the usage of DMPs tailored for Lattice Gauge Theories should be designed.
\begin{figure}[h]
\vspace{-0.25cm}
    \centering
    \includegraphics[height=4.05cm]{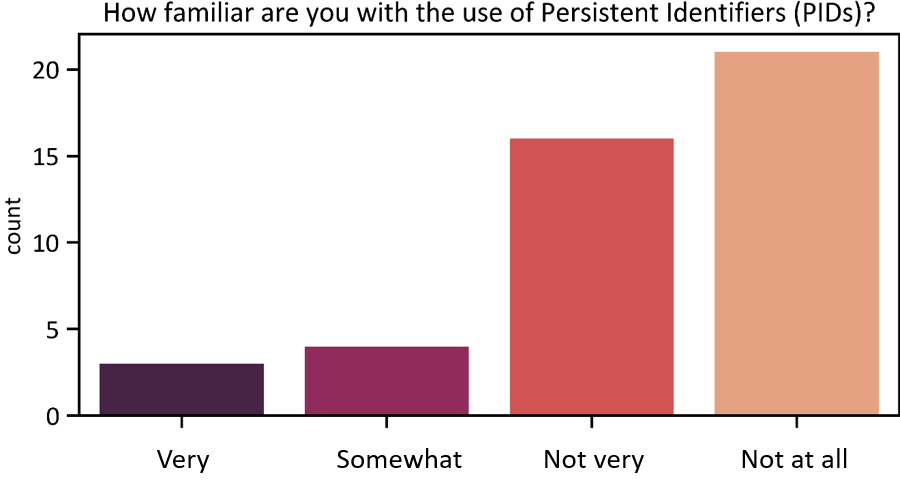}  \vspace{-0.25cm}
    \caption{ Replies to the question ``How familiar are you with the use of Persistent Identifiers?''}
    \label{fig:pid_familiarity}
    \vspace{-0.25cm}
\end{figure}

One of the important requirements to ensure findability is the assignement of a Persistent Identifier (PID) to a dataset. A PID is a long-lasting reference to a resource; unlike URLs, which may break, a PID reliably points to an entity. A commonly used PID is the DOI (Digital Object Identifier), which is a persistent identifier for things or entities such as journal articles, books, and datasets. \href{https://zenodo.org/}{Zenodo}, for example, is a repository which assigns a DOI to an uploaded dataset. Turning now to the survey's question, how familiar the participant is with the use of PIDs, the replies in Fig.~\ref{fig:pid_familiarity} indicate that most participants know very little or even nothing regarding the use of PIDs.

\begin{figure}[h]
    \vspace{-0.5cm}
    \centering
    \includegraphics[height=4.75cm]{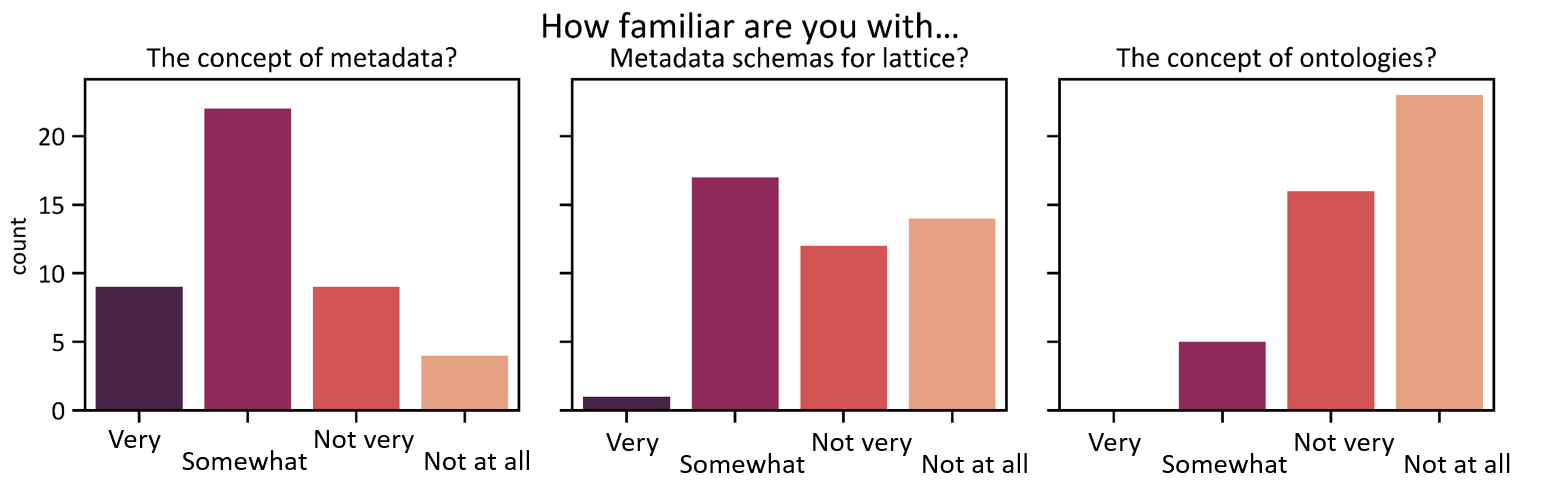}  
    \vspace{-0.25cm}
    \caption{Results on the question of how familiar participants are with 1. The concept of metadata (\underline{left panel}) 2. Metadata schemas for lattice (\underline{middle panel}) 3. The concept of ontologies (\underline{right panel}).}
    \label{fig:semantics_familiarity}
       \vspace{-0.25cm}
\end{figure}
The usage of the right semantic vocabularies (such as metadata schemas and ontologies) play an important role in the findability, interoperability and accessibility of datasets, as we explained in the previous section. Regarding ontologies---which can be thought of as a higher level semantic structure than a metadata schema---no such schemas have been developed so far in order to address interoperability between disciplines. Ontologies can indeed enhance interoperability, and could transform the interconnection of Lattice Gauge Theories with other branches of physics such as experimental High Energy Physics and Condensed Matter. Currently such cross-disciplinary connection may not be strong enough to justify the creation of such semantic structures, but this may benefit from consideration in the future. When it comes to knowledge of the Lattice Community on the above topics (Fig.~\ref{fig:semantics_familiarity}), according to the replies on the survey's question how familiar the researcher is with the concept of metadata most of participants appear to exhibit some familiarity. Turning now to familiarity with metadata schemas for lattice, most respondents appear to know very little about this topic; while ILDG has been a common metadata schema for gauge configurations for some time, this observation correlates with the fact that currently the ILDG infrastructure is infrequently referred to in publications~\cite{Bennett:2022klt} and in some cases has been inactive for some time~\cite{ILDG_Karsch}. Finally, and as expected, most participants were not familiar with the concept of ontologies.

Accessibility of data requires the existence of trustworthy repositories. A trusted repository provides a  reliable and long-term access to managed digital resources now as well as in the future. A system of metrics which can be used in order to certify a repository as trusted is \href{https://www.coretrustseal.org/}{CoreTrustSeal}. A large number of trusted repositories which can be used by individual researchers are accessible online; examples include \href{https://b2share.eudat.eu/}{B2SHARE} and \href{https://zenodo.org/}{Zenodo}. In the question what service you use to publish data/code the most frequent replies in order are GitHub, arXiv and then Zenodo. Clearly, there is a misunderstanding on what kind of repositories one should use to curate data. For instance, although arXiv is ideal for self archiving articles it cannot provide all the necessary requirements a trusted repository should meet. Furthermore, GitHub is ideal for collaborating on code, but its identifiers are not persistent; a repository must be mirrored on a service such as Zenodo to obtain a DOI. 

\begin{figure}[h]
    \vspace{-0.25cm}
    \centering
    \includegraphics[height=5.00cm]{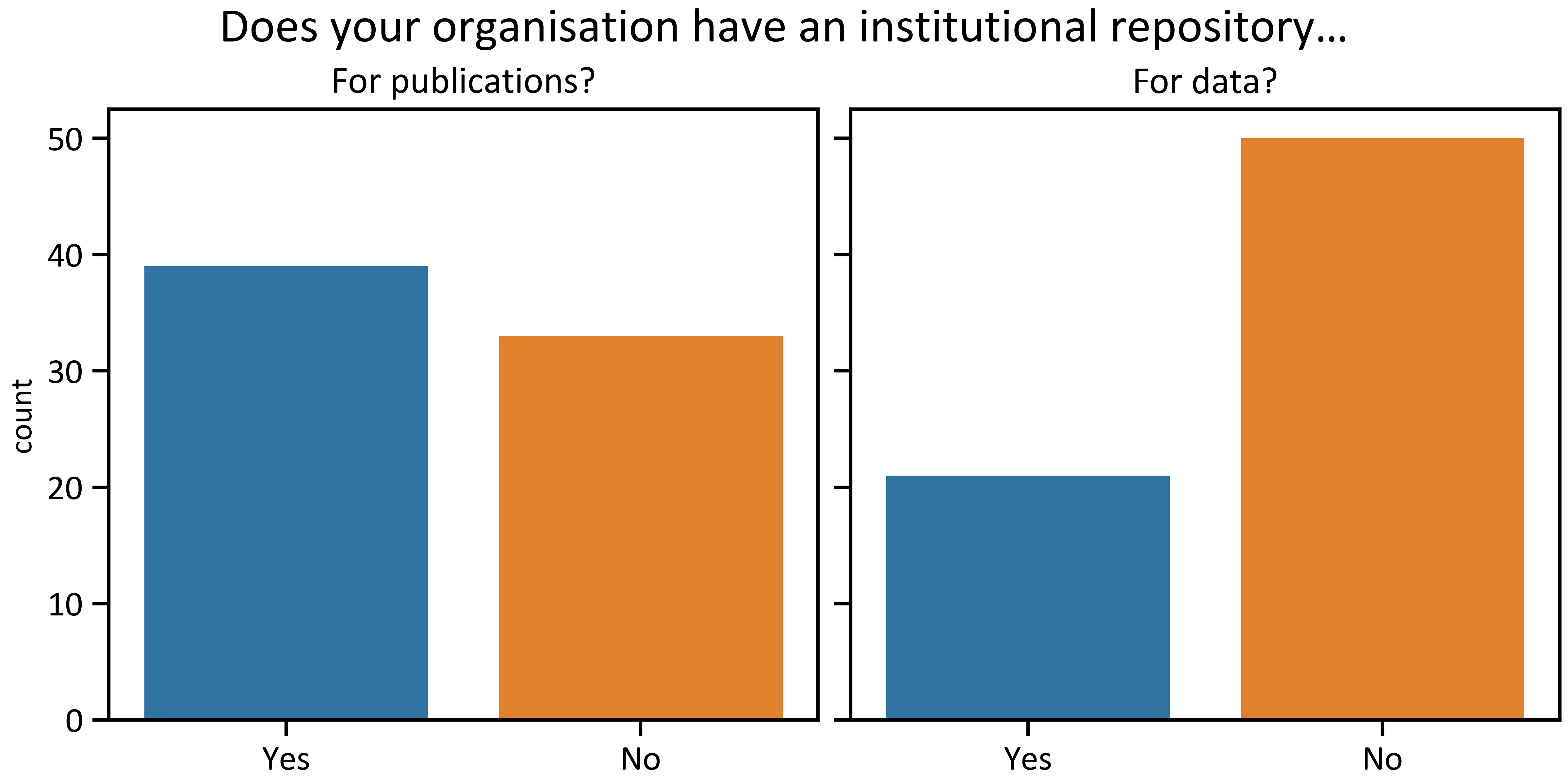}  
    \vspace{-0.25cm}
    \caption{Replies to the question ``Does your organisation have institutional repository for publications?'' (\underline{left panel}) and ``\dots for data?'' (\underline{right panel}).}
    \label{fig:institutional_repositories}
    \vspace{-0.25cm}
\end{figure}

Building capacity for Open Science could be enhanced by institutional efforts. In addition to equipping researchers with soft skills via training, organisations also provide access to institutional repositories. The replies to the question whether the participant's organisation has institutional repository for publications and data are summarised in Fig.~\ref{fig:institutional_repositories}.
Obviously, most institutions in an effort to support the concept of Open Access, provide their own researchers repositories for deposing articles. However, when it comes to sharing open data, only a few institutions appear to provide the suitable repositories, reflecting the fact that harmonisation with Open Science is not yet at a mature stage, even at institutional level.

\begin{figure}[h]
\vspace{-0.5cm}
    \centering
    \includegraphics[height=5.0cm]{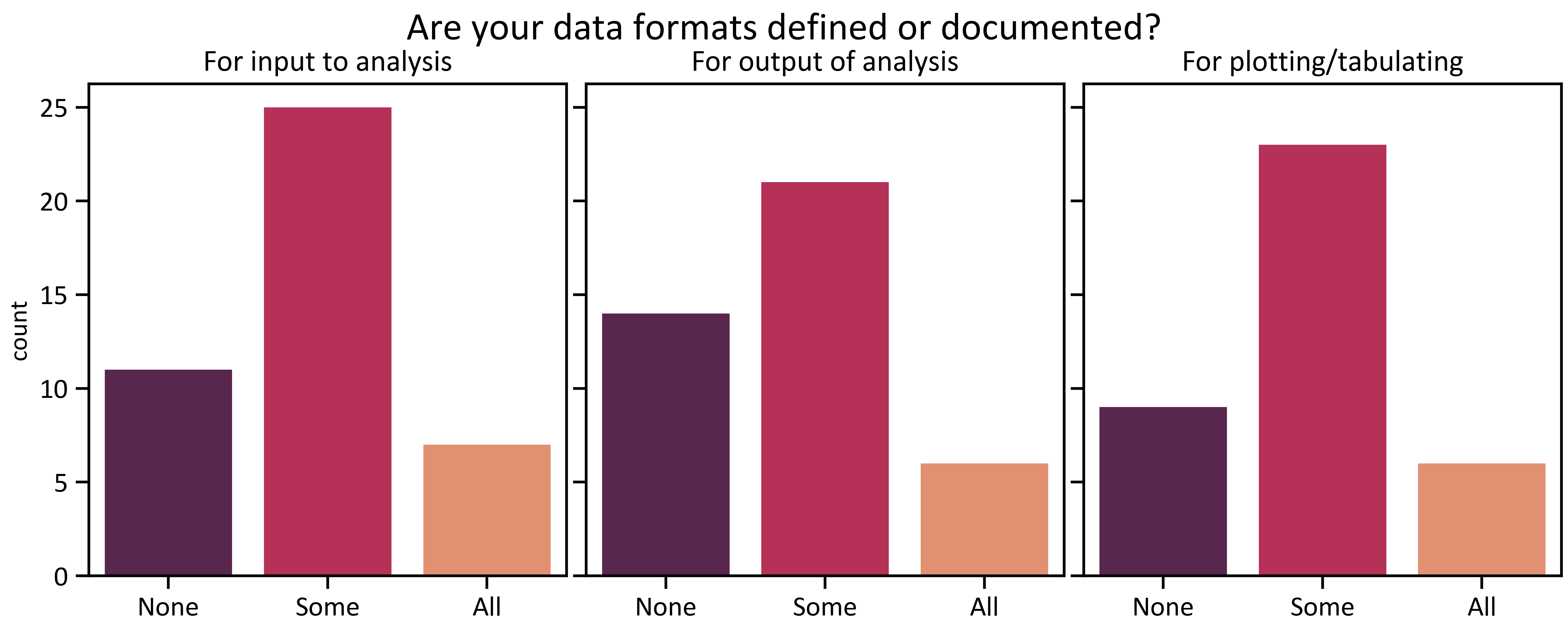}  
    \caption{The results on the question ``Are your data formats defined or documented?'' for 1. input to analysis (\underline{left}), 2. for output to analysis (\underline{middle}), for plotting/tabulating (\underline{right}).}
    \label{fig:data_formats_documented}
    \vspace{-0.25cm}
\end{figure}

Finally, the last part of this section focuses on the reproducibility as well as reusability of data. Good practises to ensure reusability include the precise definition or documentation of the used digital objects' formats. For the question whether participant's data formats are defined or documented the responses are presented in Fig.~\ref{fig:data_formats_documented}. The relevant histograms reveal that only a few researchers provide definition/documentation for all their openly accessible digital objects. When not defined nor documented, a dataset is practically unusable by anyone else except by the person who created it, and, thus, does not serve reusability. Another feature which can enable both reproducibility and reusability is the use of automated workflows. In the question if the participant ever had an automated workflow, and whether she/he would consider publishing it concurrently with the corresponding paper $\sim$65\% replied positively. This demonstrates the willingness of the lattice community to support the reusability of their results. Hence, the existence of tools which could enable lattice gauge theorists to create automatized workflows could provide an additional incentive to support such a concept.  

\begin{figure}[h]
    \vspace{-0.5cm}
    \centering
    \includegraphics[height=5.0cm]{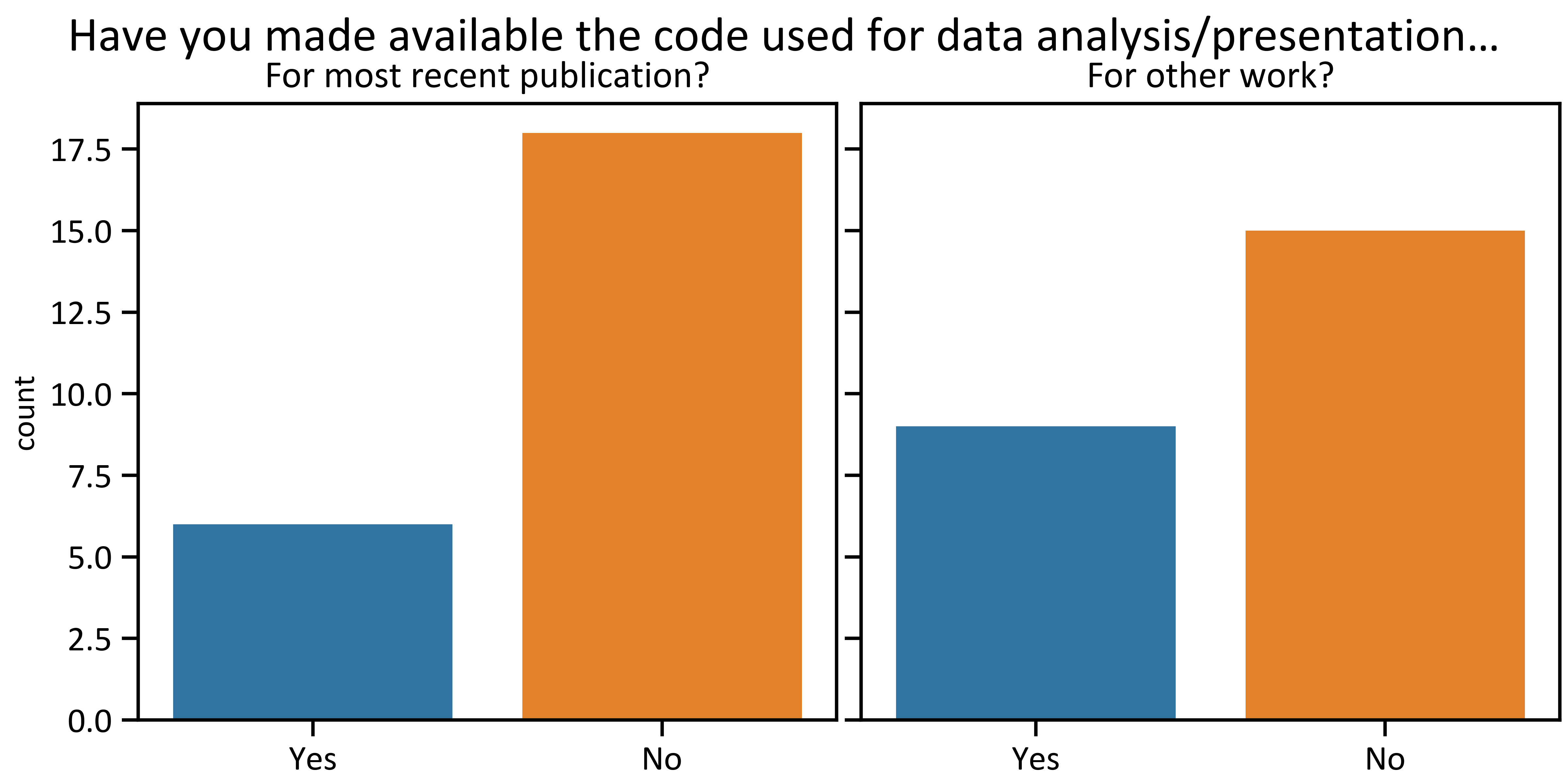}  
    \caption{The results on the question ``Have you made available the code used for data analysis/presentation for most recent publication?'' (left) and ``\dots for other work?'' (right)}
    \label{fig:made_code_available}
\end{figure}

For the question of whether the participant has ever made his/her code for data analysis or presentation openly available only $\sim$ 30 \% replied positively. When followed up as to why the they had not made their code available, most participants replied ``Full workflow is not automated, so it would not help reproducibility'' as well a ``code is not reusable in any other context''. We emphasize that even if not fully automated, the existence of a workflow can serve the reusability of a code. In addition, to be fully open, and to enable complete verification of a work where full detail could not be presented in a paper, analysis code and other software products should be published openly on the same footing as data and papers.

In the question of whether the reader has had requests to make their data/code available, most replies were positive with time of providing the data/code ranging from less than a day to a month. This demonstrates that reusability of data in lattice should be seriously considered, and having this data publicly available could minimize the delays induced by the human interaction. Lastly, when asked how often would their research benefit from access to others' data, $~\sim$ 75 \% of the participants appeared to have benefited at least once by using others' data!
\vspace{-0.25cm}
% *********************************************
% * Conclusions
% *********************************************
\section{Conclusions}
\label{sec:conclusions}
\begin{figure}[h]
\vspace{-0.6cm}
    \centering
    \includegraphics[height=4.0cm]{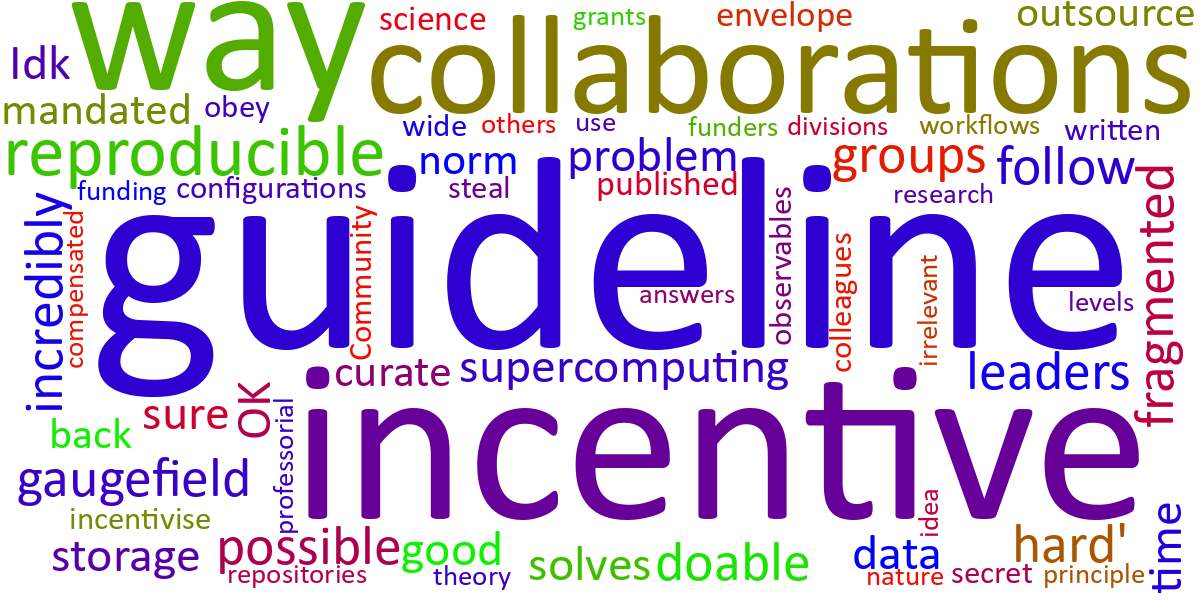}
 \hspace{0.15cm}  
    \includegraphics[height=4.25cm]{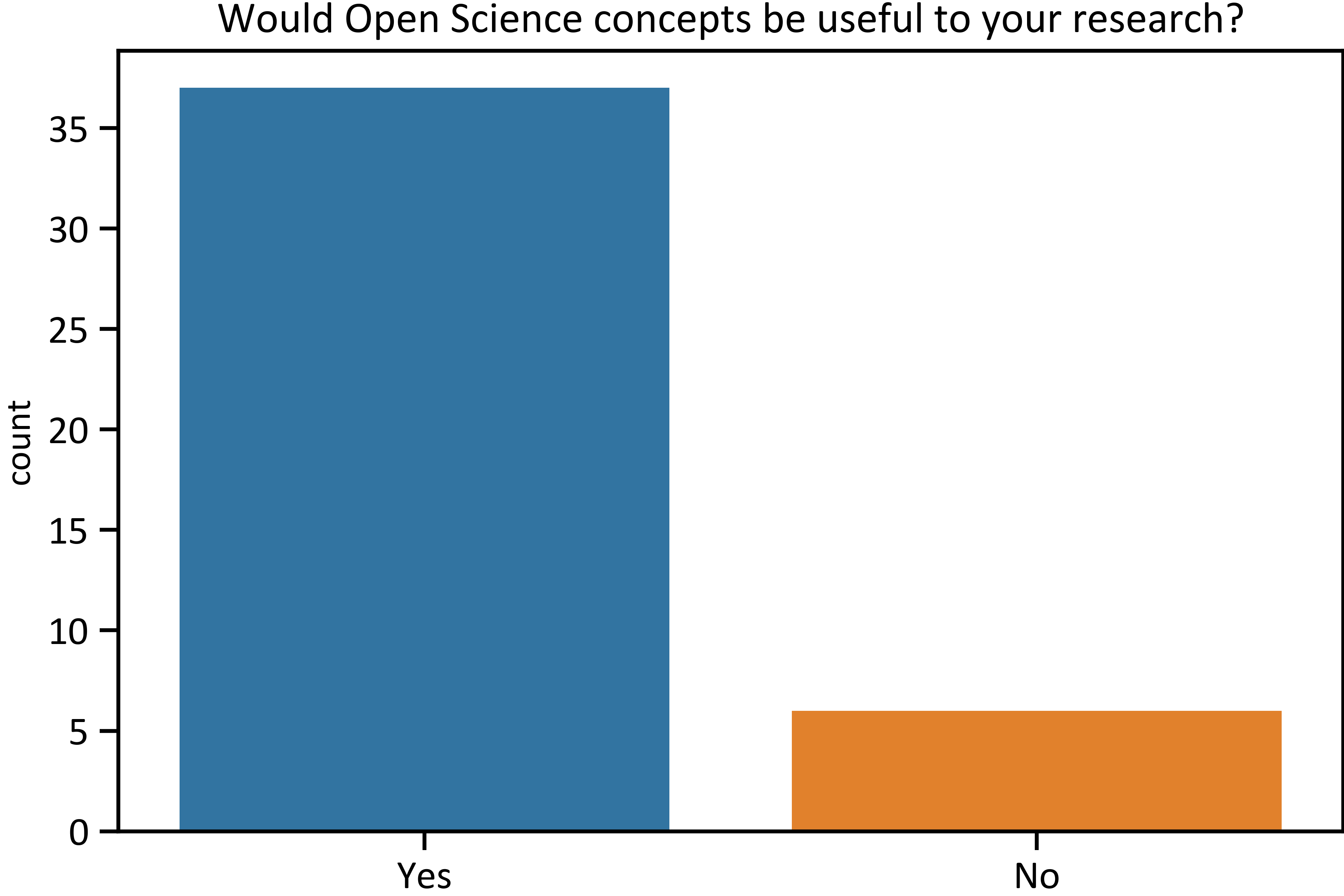}
    \caption{ \underline{Left panel:} A word cloud with the replies at the question ``What do you consider as incentives for adopting Open Science practice?'' \underline{Right panel:} The replies at the question ``Would Open Science concept be useful to your research?''}
    \label{fig:open_science_uesful}
\end{figure}

\vspace{-0.25cm}
Undoubtedly, Open Science is the new research communication {\it modus operandi} and as the survey indicates, Lattice Gauge Theorists need to catch up and harmonise the sharing of their research according to the relevant principles. This can be done by adopting Open Science best practises as well as undergoing the right training. In the question on what respondents consider as incentives for adopting Open Science, most researchers replied ``guidelines''. This manuscript provides a first draft of some general best practises the modern Lattice Gauge Theorist should adopt. It is clear that practising Open Science is a time consuming process, however the Lattice Gauge Theory community is willing to become more open with the majority of us foreseeing benefits; this can be seen in the replies on the survey's question of whether Open Science concepts are useful for one's research in Fig.~\ref{fig:open_science_uesful}. To further support the Lattice Gauge Theory community we will publish Lattice-tailored best practises in our longer write-up, self paced lectures, as well as tools which can be used towards reproducibility of data. 

\vspace{-0.05cm}

\section*{Acknowledgements}
\vspace{-0.25cm}
We would like to express our gratitude to all our lattice colleagues who have invested effort in filling out the survey and, thus, providing all the necessary information to create a landscape of the current Open Science status. A.A. and E.P. have been financially supported by NI4OS-Europe funded by the European Commission under the Horizon 2020 European research infrastructures grant agreement no.~857645. The work of E.B. and J.L. is supported by the UKRI Science and Technology Facilities Council (STFC)
Research Software Engineering Fellowship EP/V052489/1.

\bibliographystyle{apsrev}
\bibliography{biblio_NEW}

\begin{thebibliography}{7}
\expandafter\ifx\csname natexlab\endcsname\relax\def\natexlab#1{#1}\fi
\expandafter\ifx\csname bibnamefont\endcsname\relax
  \def\bibnamefont#1{#1}\fi
\expandafter\ifx\csname bibfnamefont\endcsname\relax
  \def\bibfnamefont#1{#1}\fi
\expandafter\ifx\csname citenamefont\endcsname\relax
  \def\citenamefont#1{#1}\fi
\expandafter\ifx\csname url\endcsname\relax
  \def\url#1{\texttt{#1}}\fi
\expandafter\ifx\csname urlprefix\endcsname\relax\def\urlprefix{URL }\fi
\providecommand{\bibinfo}[2]{#2}
\providecommand{\eprint}[2][]{\url{#2}}

\bibitem[{\citenamefont{Pontika and Knoth}(2015)}]{pontika2015open}
\bibinfo{author}{\bibfnamefont{N.}~\bibnamefont{Pontika}} \bibnamefont{and}
  \bibinfo{author}{\bibfnamefont{P.}~\bibnamefont{Knoth}},
  \emph{\bibinfo{title}{Open science taxonomy}} (\bibinfo{year}{2015}),
  \urlprefix\url{https://doi.org/10.6084/m9.figshare.1508606.v3}.

\bibitem[{\citenamefont{Commission et~al.}(2021)\citenamefont{Commission, for
  Research, and Innovation}}]{doi/10.2777/69533}
\bibinfo{author}{\bibfnamefont{E.}~\bibnamefont{Commission}},
  \bibinfo{author}{\bibfnamefont{D.-G.} \bibnamefont{for Research}},
  \bibnamefont{and} \bibinfo{author}{\bibnamefont{Innovation}},
  \emph{\bibinfo{title}{Horizon Europe : Horizon Europe : apply for your
  funding now!}} (\bibinfo{publisher}{Publications Office of the European
  Union}, \bibinfo{year}{2021}).

\bibitem[{\citenamefont{Commission et~al.}(2016)\citenamefont{Commission, for
  Research, and Innovation}}]{doi/10.2777/940154}
\bibinfo{author}{\bibfnamefont{E.}~\bibnamefont{Commission}},
  \bibinfo{author}{\bibfnamefont{D.-G.} \bibnamefont{for Research}},
  \bibnamefont{and} \bibinfo{author}{\bibnamefont{Innovation}},
  \emph{\bibinfo{title}{Realising the European open science cloud : first
  report and recommendations of the Commission high level expert group on the
  European open science cloud}} (\bibinfo{publisher}{Publications Office},
  \bibinfo{year}{2016}).

\bibitem[{\citenamefont{Wilkinson et~al.}(2016)\citenamefont{Wilkinson,
  Dumontier, Aalbersberg, Appleton, Axton, Baak, Blomberg, Boiten,
  da~Silva~Santos, Bourne et~al.}}]{wilkinson2016fair}
\bibinfo{author}{\bibfnamefont{M.~D.} \bibnamefont{Wilkinson}},
  \bibinfo{author}{\bibfnamefont{M.}~\bibnamefont{Dumontier}},
  \bibinfo{author}{\bibfnamefont{I.~J.} \bibnamefont{Aalbersberg}},
  \bibinfo{author}{\bibfnamefont{G.}~\bibnamefont{Appleton}},
  \bibinfo{author}{\bibfnamefont{M.}~\bibnamefont{Axton}},
  \bibinfo{author}{\bibfnamefont{A.}~\bibnamefont{Baak}},
  \bibinfo{author}{\bibfnamefont{N.}~\bibnamefont{Blomberg}},
  \bibinfo{author}{\bibfnamefont{J.-W.} \bibnamefont{Boiten}},
  \bibinfo{author}{\bibfnamefont{L.~B.} \bibnamefont{da~Silva~Santos}},
  \bibinfo{author}{\bibfnamefont{P.~E.} \bibnamefont{Bourne}},
  \bibnamefont{et~al.}, \bibinfo{journal}{Scientific data}
  \textbf{\bibinfo{volume}{3}} (\bibinfo{year}{2016}).

\bibitem[{\citenamefont{Karsch}()}]{ILDG_Karsch}
\bibinfo{author}{\bibfnamefont{F.}~\bibnamefont{Karsch}},
  \emph{\bibinfo{title}{Update on the international lattice data grid}},
  \bibinfo{howpublished}{\url{https://indico.hiskp.uni-bonn.de/event/40/contributions/852/attachments/425/696/ILDG-talk.pdf}}.

\bibitem[{\citenamefont{Athenodorou et~al.}(2022)\citenamefont{Athenodorou,
  Bennett, Lenz, and Papadopoulou}}]{andreas_athenodorou_2022_6980070}
\bibinfo{author}{\bibfnamefont{A.}~\bibnamefont{Athenodorou}},
  \bibinfo{author}{\bibfnamefont{E.}~\bibnamefont{Bennett}},
  \bibinfo{author}{\bibfnamefont{J.}~\bibnamefont{Lenz}}, \bibnamefont{and}
  \bibinfo{author}{\bibfnamefont{E.}~\bibnamefont{Papadopoulou}},
  \emph{\bibinfo{title}{{Survey on lattice data analysis, presentation, and
  curation practices}}} (\bibinfo{year}{2022}),
  \urlprefix\url{https://doi.org/10.5281/zenodo.6980070}.

\bibitem[{\citenamefont{Bennett}(2022)}]{Bennett:2022klt}
\bibinfo{author}{\bibfnamefont{E.}~\bibnamefont{Bennett}}, in
  \emph{\bibinfo{booktitle}{{39th International Symposium on Lattice Field
  Theory}}} (\bibinfo{year}{2022}), \eprint{2211.15547}.

\end{thebibliography}

\end{document}